\documentclass[prl,twocolumn,showpacs,preprintnumbers,amsmath,amssymb]{revtex4}
\usepackage{graphicx}
\usepackage{dcolumn}
\usepackage{bm}
\begin{document}
%\preprint{IOPB}
\title{\bf A new microscopic nucleon-nucleon interaction derived from\\ relativistic mean field theory}

\author{BirBikram Singh}
\author{M. Bhuyan}
\author{S. K. Patra}\email{patra@iopb.res.in}
%\address[Singh]{Institute of Physics, Sachivalaya Marg, Bhubaneswar 751005, India}

\affiliation{Institute of Physics, Sachivalaya Marg, Bhubaneswar - 751 005, India}
\author{Raj K. Gupta}
\affiliation{Department of Physics, Panjab University, Chandigarh 160014, India.}

\begin{abstract}
A new microscopic nucleon-nucleon (NN) interaction has been derived for the first time from the popular relativistic mean 
field theory (RMFT) Lagrangian. The NN interaction so obtained remarkably relate to the inbuilt fundamental parameters of 
RMFT. Furthermore, by folding it with the RMFT-densities of cluster and daughter nuclei to obtain the optical potential, 
it's application is also examined to study the exotic cluster radioactive decays, and results obtained found comparable 
with the successfully used M3Y phenomenological effective NN interactions. The presently derived NN-interaction can also 
be used to calculate a number of other nuclear observables.
\end{abstract}

\maketitle
%\keywords{M3Y effective nucleon-nucleon interaction; Relativistic mean field theory}
%%\vfil
%%\eject
%%\baselineskip 20pt

Nucleon-nucleon (NN) interaction is an active area of research since the discovery of neutron, decades back. NN interaction 
was conceived to be mediated by mesons, much before their discovery. Though substantial progress has taken place to 
understand it in a number of theoretical (and experimental) attempts, so far it has remained an open problem. Large number 
of interactions have been constructed via studying NN scattering. But extensive modification in the scattering behaviour 
due to the presence of many other nucleons inside the nucleus make it appropriate to use the phenomenological effective or 
averaged interactions instead, which typically depend on the local density of nuclear matter. The nucleus-nucleus 
potentials obtained by using effective NN interactions are used to study the number of observed nuclear phenomena and 
hence also provide a useful understanding of the NN interaction. For example, the effective NN interaction has been 
remarkably related to the nucleus-nucleus potential in the double folding model (DFM) \cite{sat79}. 

The microscopic heavy-ion scattering potential of interest is obtained in DFM \cite{sat79} by using an effective 
nucleon-nucleon (NN) interaction, like the M3Y plus a zero-range pseudo-potential or a density-dependent M3Y (DDM3Y), 
folded over the matter densities of the interacting nuclei. It is relevant to mention here that the simplified spin- and 
isospin-independent (S=T=0) M3Y effective NN interaction \cite{sat79} has been used widely and successfully in a number of 
applications (see, e.g., \cite{gold83,krishi10,singh10}). Actually, an effective NN interaction is S (and T)-dependent 
\cite{love70,love72}, and generally carries three components as
\begin{equation}
v_{eff}=V^{C}(r)+V^{LS}(r)\vec{L}.\vec{S}+V^{T}(r)\hat{S}_{12},
\label{eq:1}
\end{equation}
where $r$ is the relative distance and $\vec{L}.\vec{S}$ and $\hat{S}_{12}$ are the usual spin-orbit and tensor operators, 
respectively. The central component \cite{love70} is
\begin{equation}
v_{eff}^{C}=V_{0}(r)+V_{\sigma}(r){\sigma}_{1}.{\sigma}_{2}+V_{\tau}(r){\tau}_{1}.{\tau}_{2}
+V_{{\sigma}{\tau}}(r)({\sigma}_{1}.{\sigma}_{2})({\tau}_{1}.{\tau}_{2}),
\label{eq:2}
\end{equation}
with radial and spin-, isospin-, spin-isospin-dependent parts, respectively. In principle, Quantum Chromodynamics (QCD) 
may be used to obtain the radial dependence of these components, instead, these are normally expressed in terms of Yukawa's 
or other functional fits to the experimental data. Hence, the available NN potentials are phenomenological in nature and 
a complete microscopic NN potential derived from QCD is yet to be achieved. 

In this Letter, we have derived the microscopic NN interaction from the linear relativistic mean field theory (RMFT) 
\cite{horo81,ser86,rein89,patra91} Lagrangian, rather than using a simple phenomenological prescription. It is relevant to 
mention here that the RMFT is an established approach for the accurate description of nuclear bulk properties (the binding 
energy, root-mean-square radii and clustering) over the entire region of nuclear chart including the superheavy nuclei. 
This has also been applied successfully to infinite nuclear matter, like the equation of state EOS and neutron star 
\cite{arum104}. The NN interaction derived in such a manner could be used to obtain the double folding potential, which 
can have crucial significance for futher understanding of the NN interaction, a number of nuclear properties, as well as 
for the experiments to be planned in this direction. We employ it here to study the exotic cluster radioactive decays and 
compare our results with that of the use of phenomenological M3Y effective NN interaction.
\begin{figure*}
\vspace{-1.8cm}
\includegraphics[width=1.60\columnwidth, clip=true]{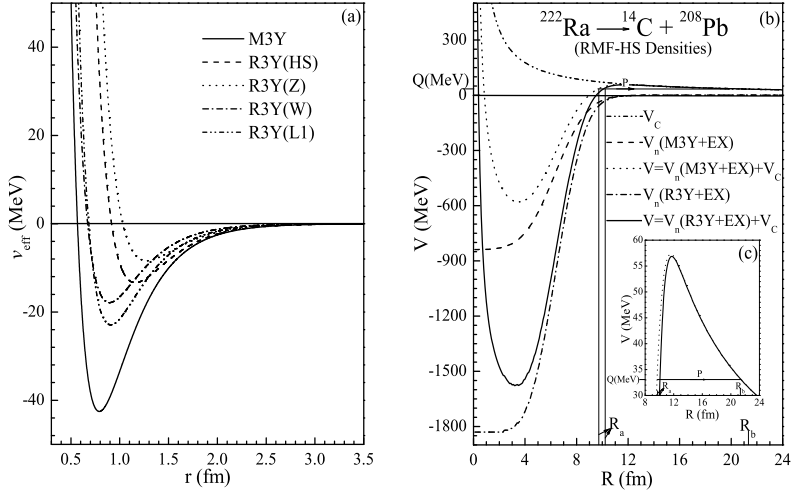}
\vspace{-2.8cm}
\caption{(a) R3Y (for different parameter sets of Table I) and M3Y effective NN interaction potential as a function of $r$. 
(b) The total nucleus-nucleus optical potential $V(R)$ and the individual contributions [$V_n(R)(M3Y+EX)$ and 
$V_n(R)(R3Y+EX)$ for HS parameter set, and the Coulomb $V_C(R)$] as a function of radial separation R.
(c) The inset of (b); same as (b) but with a changed scale in order to magnify the barrier height and position. 
}
\end{figure*}

The linear, relativistic mean field Lagrangian density for a nucleon-meson many-body system 
\cite{horo81,ser86,rein89,patra91} is
%% Equation 1
%\begin{widetext}
\begin{eqnarray}
&&L=\overline{\psi_{i}}\{i\gamma^{\mu}\partial_{\mu}-M\}\psi_{i}+{\frac12}\partial^{\mu}
\sigma\partial_{\mu}\sigma-{\frac12}m_{\sigma}^{2}\sigma^{2}-g_{\sigma}\overline{\psi_{i}}\psi_{i}
\sigma\nonumber\\&&-{\frac14}\Omega^{\mu\nu}\Omega_{\mu\nu}
+{\frac12}m_{w}^{2}V^{\mu} V_{\mu}-g_{w}\overline\psi_{i}\gamma^{\mu} \psi_{i}V_{\mu}-{\frac14}
\vec{B}^{\mu\nu}.\vec{B}_{\mu\nu}\nonumber\\
&&+{\frac12}m_{\rho}^{2}{\vec R^{\mu}} .{\vec{R}_{\mu}}
%\nonumber\\
%&&
-g_{\rho}\overline\psi_{i}\gamma^{\mu}\vec{\tau}\psi_{i}.\vec{R^{\mu}}-{\frac12}m_{\delta}^{2}
\delta^{2}+g_{\delta}\overline\psi_{i}\delta\vec{\tau}\psi_{i},
\end{eqnarray}
%\end{widetext}
%
where, the field for $\sigma$ meson is denoted by $\sigma$, that for $\omega$ meson by $V_{\mu}$, and that for the 
iso-vector $\rho$ and $\delta$ mesons by $\vec{R}_{\mu}$ and $\delta$, respectively. $A_{\mu}$ denotes the electromagnetic 
field. The $\psi_{i}$ are the Dirac spinors for the nucleons. An iso-spin is denoted by $\tau$. Here $g_{\sigma}$, 
$g_{\omega}$, $g_{\rho}$ and $g_{\delta}$ are the coupling constants for $\sigma$, $\omega$, $\rho$ and $\delta$ mesons, 
respectively. M, $m_{\sigma}$, $m_{\omega}$, $m_{\rho}$ and $m_{\delta}$ are the masses of the nucleons, $\sigma$, 
$\omega$, $\rho$ and $\delta$ mesons, respectively. $\Omega^{\mu\nu}$ and $\vec{B}_{\mu\nu}$ are the field tensors for the 
$V^{\mu}$ and $\vec{R}_{\mu}$, respectively. In this Langrangian the contribution of $\pi$ meson has not been taken in to 
account as, at the mean-field level, its contribution is zero due to its pseudoscalar nature \cite{ser86,brock78}.

From the above relativistic Lagrangian, we obtain the field equations for the nucleons and mesons as,
\begin{eqnarray}
\Bigl(-i\alpha.\bigtriangledown+\beta(M+g_{\sigma}\sigma)+g_{\omega}\omega+\nonumber\\
g_{\rho}{\tau}_3{\rho}_3+g_{\delta}\delta{\tau}\Bigr){\psi}_i={\epsilon}_i{\psi}_i,\\
(-\bigtriangledown^{2}+m_{\sigma}^{2})\sigma(r)=-g_{\sigma}{\rho}_s(r),\\
(-\bigtriangledown^{2}+m_{\omega}^{2})V(r)=g_{\omega}{\rho}(r),\\
(-\bigtriangledown^{2}+m_{\rho}^{2})\rho(r)=g_{\rho}{\rho}_3(r),\\
(-\bigtriangledown^{2}+m_{\delta}^{2})\delta(r)=-g_{\delta}{\rho}_3(r),
\end{eqnarray}
respectively, for Dirac nucleons, $\sigma$, $\omega$, $\rho$, $\delta$ mesons.

In the limit of one-meson exchange, for a heavy and static baryonic medium, the solution of single nucleon-nucleon 
potential for scalar ($\sigma$, $\delta$) and vector ($\omega$, $\rho$) fields are
$V_{\sigma}(r)=
-\frac{g_{\sigma}^{2}}{4{\pi}}\frac{e^{-m_{\sigma}r}}{r}$,
$V_{\delta}(r)=
-\frac{g_{\delta}^{2}}{4{\pi}}\frac{e^{-m_{\delta}r}}{r}$ and
$V_{\omega}(r)=
\frac{g_{\omega}^{2}}{4{\pi}}\frac{e^{-m_{\omega}r}}{r}$,
 $V_{\rho}(r)=
+\frac{g_{\rho}^{2}}{4{\pi}}\frac{e^{-m_{\rho}r}}{r}$, respectively.
The resultant effective nucleon-nucleon interaction, obtained from the summation of the scalar and vector parts of the 
single meson fields, is defined as \cite{brock78,miller72,brock77}
%\begin{equation}
\begin{eqnarray}
&& v_{eff}(r)=V_{\omega}+V_{\rho}+V_{\sigma}+V_{\delta}
\nonumber \\
&&
=\frac{g_{\omega}^{2}}{4{\pi}}\frac{e^{-m_{\omega}r}}{r}+
\frac{g_{\rho}^{2}}{4{\pi}}\frac{e^{-m_{\rho}r}}{r}
-\frac{g_{\sigma}^{2}}{4{\pi}}\frac{e^{-m_{\sigma}r}}{r}-\frac{g_{\delta}^{2}}{4{\pi}}
\frac{e^{-m_{\delta}r}}{r}.
\label{eq:9}
\end{eqnarray}
%\end{equation}
For a normal nuclear medium, the contribution $V_{\delta}$ of $\delta$-meson can be neglected, compared to the magnitudes 
of both $V_{\omega}$ and $V_{\sigma}$. Hence, Eq. (9) becomes
\begin{equation}
v_{eff}(r)=\frac{g_{\omega}^{2}}{4{\pi}}\frac{e^{-m_{\omega}r}}{r}+
\frac{g_{\rho}^{2}}{4{\pi}}\frac{e^{-m_{\rho}r}}{r}-
\frac{g_{\sigma}^{2}}{4{\pi}}\frac{e^{-m_{\sigma}r}}{r},
\label{eq:10}
\end{equation}
where, the values of $\frac{g_{\omega}^2}{\pi}$, $\frac{g_{\rho}^2}{\pi}$ and $\frac{g_{\sigma}^2}{\pi}$ are listed in 
Table I for different parameter sets of RMF models \cite{horo81,rein89,theis83}, except for W and L1 sets for which only 
$\frac{g_{\omega}^2}{\pi}$ and $\frac{g_{\sigma}^2}{\pi}$ are given since contribution of $\rho$ meson is ignored for 
these parameter sets.

\begin{table*}
\tabcolsep 0.3cm
\caption{The values of $m_{\sigma}$, $m_{\omega}$, $m_{\rho}$  (in MeV) and $g_{\sigma}$, $g_{\omega}$, $g_{\rho}$ 
for different RMF works, along with $\frac{g_{\sigma}^2}{\pi}$,
$\frac{g_{\omega}^2}{\pi}$, $\frac{g_{\rho}^2}{\pi}$ (in MeV).}
\label{defparagcl}
\begin{center}
\begin{tabular}{l l l l l l l l l l}
\hline
Set&$m_{\sigma}$& $m_{\omega}$ & $m_{\rho}$& $g_{\sigma}$ &$g_{\omega}$&$g_{\rho}$& $\frac{g_{\sigma}^2}{\pi}$&
$\frac{g_{\omega}^2}{\pi}$&$\frac{g_{\rho}^2}{\pi}$\\ \hline
HS \cite{horo81}       & 520 & 783&770 & 10.47 &13.80 & 08.08 &6882.64&11956.94&4099.06 \\
Z  \cite{rein89}     & 551.31 & 780 & 763 & 11.19 & 13.83 & 10.89 &7861.80&12008.98&7445.91 \\
W \cite{rein89}     & 550 & 783 &$-$& 09.57 & 11.67 &$-$&5750.24&8550.74&$-$ \\
L1 \cite{rein89}          & 550 & 783 &$-$& 10.30 & 12.60 &$-$&6660.95&9967.88&$-$ \\
\hline
\end{tabular}
\end{center}
\end{table*}

Using the HS parameters of Table I in Eq. (10), we get
\begin{equation}
v_{eff}(r)= 11956\frac{e^{-3.97r}}{4r}+4099\frac{e^{-3.90r}}{4r}-
6882\frac{e^{-2.64r}}{4r},
\label{eq:11}
\end{equation}
and for the L1 parameters, Eq. (10) becomes
\begin{equation}
v_{eff}(r)= 9967\frac{e^{-3.97r}}{4r}-6660\frac{e^{-2.79r}}{4r},
\label{eq:12}
\end{equation}
with the corresponding effective NN-potentials as shown in Fig. 1(a). 

On the other hand, the M3Y effective interaction, obtained from a fit of the G-matrix elements based on Reid-Elliott 
soft-core NN interaction \cite{sat79}, in an oscillator basis, is the sum of three Yukawa's (M3Y) with ranges 0.25 fm for 
a medium-range attractive part, 0.4 fm for a short-range repulsive part and 1.414 fm to ensure a long-range tail of the 
one-pion exchange potential (OPEP). The widely used M3Y effective interaction $v_{eff}(r)$ is given by
\begin{equation}
v_{eff}(r)=7999\frac{e^{-4r}}{4r}-2134\frac{e^{-2.5r}}{2.5r},
\label{eq:13}
\end{equation}
where ranges are in fm and the strength in MeV. Note that Eq. (13) represents the spin- and isospin-independent parts 
of the central component of the effective NN interaction [Eqs. (1) and (2)], and that the OPEP contribution is absent here.
Comparing Eqs. (11) and/ (12) and (13), we find similarity in the behaviour of the NN-interaction and feel that equation 
(10) can be used to obtain the nucleus-nucleus optical potential.

Now we demonstrate the application of equations (10) and (13) to various nuclear systems for evaluating some of the 
physical observables in the phenomenon of exotic cluster radioactivity (CR). Fig. 1(a) illustrates the comparison between 
the M3Y effective NN interaction (Eq. (13), solid line) and its equivalent, the RMF based 3 Yukawa's (denoted R3Y) 
presentations of Eq. (10) which is based on RMF considerations involving the coupling constants $g_{\omega}$, $g_{\rho}$, 
$g_{\sigma}$ and the meson masses $m_{\omega}$,  $m_{\rho}$, $m_{\sigma}$.

Using the preformed cluster model (PCM) of Gupta and collaborators \cite{malik89,gupta94}, we deduce empirically the 
cluster preformation probability ${P_0}^{c(emp)}$ from experimental data on a few exotic cluster radioactive (CR) decays 
in the trans-lead region having doubly magic $^{208}$Pb as daughters, using the HS parameter set based spherical, 
relativistic mean field (RMF-HS) densities. It is relevant to mention here that the mass and charge densities calculated 
by using the RMF theory by some of us and collaborators \cite{patra07}, support the clustering effects in various heavy 
parents with observed cluster decays. In PCM, the decay constant $\lambda$ or half-life time $T_{1/2}$ is defined as 
\cite{singh10,malik89,gupta94} 
\begin{equation}
\lambda_{PCM}=\frac{\ln 2}{T_{1/2}}=\nu_0 P_0 P,
\label{eq:14}
\end{equation} 
with the assault frequency $\nu_0\sim 10^{21}$ $s^{-1}$ for all the cluster-decays \cite{gupta94}. An empirical estimate 
of the pre-formation factor $P_0$ can be obtained as \cite{singh10}
\begin{equation}
{P_0}^{emp}=\frac{\lambda_{Expt}}{\nu_0 P}, 
\label{eq:15}
\end{equation} 
from the experimental $\lambda_{Expt}$ values \cite{gupta94} and calculated ${\nu_0 P}$. In the following, the values of 
${P_0}^{c(emp)}$ deduced by using the R3Y and M3Y NN interactions are compared. 

\begin{table*}
\tabcolsep 0.3cm
\caption{$P$ and ${P_0}^{c(emp)}$ for cluster-decays of some parents with $^{208}$Pb as the daughter nucleus, calculated 
for the R3Y+EX  and compared with the M3Y+EX NN interaction potential, for RMF-HS densities. The experimental data on 
cluster-decay constant ${\lambda}^{c}_{Expt}$ are from \cite{gupta94}, and the Q-values are calculated by using the 
experimental ground-state binding energies \cite{audi03}.}
\label{defparagc2}
\begin{center}
\begin{small}
\begin{tabular}{l l l l l l l l}
\hline
%Parent&Cluster&   Q&   $P$&   $P$&${\lambda}^c_{Expt}$&${P_0}^{c(emp)}$&${P_0}^{c(emp)}$\\
Parent&Cluster&   Q&\multicolumn{2}{c}{$P$}   &${\lambda}^c_{Expt}$&\multicolumn{2}{c}{${P_0}^{c(emp)}$} \\
&&(MeV)&$(M3Y+EX)$&$(R3Y+EX)$&$(s^{-1})$&$(M3Y+EX)$&$(R3Y+EX)$\\ \hline
$^{222}$Ra&$^{14}$C &33.050 &$1.728\times10^{-25}$ &$2.277\times10^{-24}$ &$6.749\times10^{-12}$ 
&$1.044\times10^{-08}$ &$7.921\times10^{-10}$ \\
$^{230}$U &$^{22}$Ne&61.388 &$1.378\times10^{-29}$ &$7.615\times10^{-27}$ &$4.243\times10^{-19}$ 
&$7.664\times10^{-12}$ &$1.387\times10^{-14}$ \\
$^{231}$Pa&$^{23}$F &51.844 &$6.613\times10^{-33}$ &$1.593\times10^{-30}$ &$1.682
\times10^{-25}$ &$7.062\times10^{-15}$ &$2.932\times10^{-17}$ \\
$^{232}$U &$^{24}$Ne&62.311 &$1.047\times10^{-28}$ &$1.753\times10^{-26}$ &$2.720\times10^{-21}$ 
&$6.731\times10^{-15}$ &$4.019\times10^{-17}$ \\
$^{236}$Pu&$^{28}$Mg&79.670 &$5.710\times10^{-27}$ &$3.815\times10^{-23}$ &$1.469
\times10^{-22}$ &$6.401\times10^{-18}$ &$9.580\times10^{-22}$ \\
$^{238}$Pu&$^{30}$Mg&76.824 &$1.873\times10^{-30}$ &$1.185\times10^{-25}$ &$1.412
\times10^{-26}$ &$1.984\times10^{-18}$ &$3.136\times10^{-23}$ \\
\hline
\end{tabular}
\end{small}
\end{center}
\end{table*}

The nuclear interaction potential, $V_n(R)$, between the cluster (c) and daughter (d) nuclei, with the respective RMF-HS 
calculated nuclear matter densities $\rho_c$ and $\rho_d$, is
\begin{equation}
V_{n}(\vec{R})=\int\rho_{c}(\vec{r}_c)\rho_{d}(\vec{r}_d)v_{eff}(|\vec{r}_c-\vec{r}_d+
\vec{R}|{\equiv}r)d^{3}r_cd^{3}r_d,
\label{eq:16}
\end{equation}
obtained by using the well known double folding procedure \cite{sat79} to the M3Y (or R3Y, proposed in the present study) 
interaction, supplemented by zero-range pseudo-potential representing the single-nucleon exchange effects (EX). Adding 
Coulomb potential $V_C(R)$ (=$Z_dZ_ce^2/R$) results in cluster-daughter interaction potential $V(R)$ [$=V_n(R)+V_C(R)$], 
used for calculating the WKB penetrability P, representing relative motion R. The other details of the methodology 
followed are given in Ref. \cite{singh10}.

\begin{figure}
\vspace{-0.4cm}
\includegraphics[angle=0,width=0.80\columnwidth, clip=true]{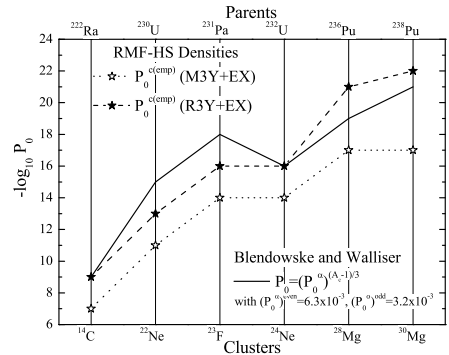}
\vspace{-0.6cm}
\caption{The ${P_0}^{c(emp)}$ for the cluster-decays, respectively, from various parents evaluated with the use of R3Y and 
M3Y  effective NN interaction compared with the phenomenological model of Blendowske-Walliser.}
\end{figure}

Fig. 1(b) illustrates the total interaction potentials $V(R)$ for $^{14}$C decay of $^{222}$Ra, obtained for both the 
M3Y+EX and R3Y+EX NN interactions using RMF-HS densities. The penetration path with an energy equal to the Q-value of decay 
is also shown here. Note that, compared to the M3Y NN interaction, the barrier is a bit lowered for the R3Y case (shown 
more clearly in the inset Fig. 1(c)) and hence $P$ increased by a few orders, as is shown in Table II for some decays. 
Consequently, the deduced ${P_0}^{c(emp)}(R3Y+EX)$ are also affected. However, interestingly, in Fig. 2 we find that the 
values of ${P_0}^{c(emp)}(R3Y+EX)$ are closer to the well accepted phenomenological formula of Blendowske-Walliser (BW) 
\cite{blend88} whereas the same for ${P_0}^{c(emp)}(M3Y+EX)$ are within two to three orders of magnitude with the BW 
results. Evidently, the effective NN interaction obtained from the RMF Lagrangian, the R3Y, is applicable to study the 
exotic cluster radioactive decays within a satisfactory precision. However, the present study has been carried out by 
taking in to consideration only the linear terms of the $\sigma$, $\omega$ and $\rho$ meson fields. Apparently, it is 
relevant as well as interesting to study the link between the RMF phenomenology and the effective NN interaction with the  
further inclusion of the non-linear terms of these fields.

Concluding, for the first time to our knowledge, we have shown in this Letter that the effective nucleon-nucleon 
interaction, here called R3Y, could be derived from the simple linear Walecka Lagragian, rather than using the simple 
phenomenological prescription. It is presented eloquently in terms of the well known inbuilt RMFT parameters of 
$\sigma$, $\omega$ and $\rho$ meson fields, i.e.,  their masses ($m_{\sigma}$, $m_{\omega}$, $m_{\rho}$) and coupling 
constants ($g_{\sigma}$, $g_{\omega}$,  $g_{\rho}$). Thus, the phenomenological M3Y NN interaction can be replaced by the 
presently derived R3Y interaction(s) for most of the calculations of nuclear observables. Moreover, we have generated here 
a bridge between R3Y and M3Y for the nucleus-nucleus folding optical potential which can be considered as a unification of 
the RMF model to predict the nuclear cluster activities, so that we can explain the cluster decay properties of the 
clustering nuclei using the R3Y instead of the M3Y potential. The improvement of R3Y interaction for use of the most 
successful non-linear RMF or E-RMF Lagrangian is straight forward. Furthermore, the present findings could be considered as 
the motivation for other similar models for the generation of different types of NN-interactions as well as the additional 
feather to RMFT for its being considered as a unified formalism to study a number of nuclear phenomena and, above all, one 
more step forward to understand the NN interaction within a well established theoretical formalism, the RMFT.

%%%%%%%%%%%%%%%%%%%%%%%%%%%%%%%%%%%%%%%%%%%%%%%%%%%%%%%%%%%%%%%%%%%%%%%%%%
\end{document}